\documentstyle{article}
\font\sqi=cmssq8
\def\DR{\rm I\kern-1.45pt\rm R}
\def\DC{\kern2pt {\hbox{\sqi I}}\kern-4.2pt\rm C}
\begin{document}
\begin{center}
{\large\bf $D-$dimensional
massless particle with extended gauge invariance}\\
\vspace{0.6cm}
{\large A. Nersessian}\\
\smallskip
{\sl Laboratory of Theoretical Physics, JINR,
 Dubna, 141980, Russia  \\
\smallskip
 Yerevan State University, A.Manoukian St., 3, Yerevan, 375025, Armenia}
             \end{center}

\begin{abstract}
 We propose the model  of $D-$dimensional massless particle
whose Lagrangian is given by  the  $N-$th extrinsic  curvature of
world-line.
The  system has $N+1$  gauge degrees of freedom constituting
$W-$like algebra; the  classical trajectories
of the model  are  space-like curves which obey the conditions
 $k_{N+a}=k_{N-a}$, $k_{2N}=0$, $a=1,\ldots,N-1$, $N\leq[(D-2)/2]$,
while the first $N$ curvatures $k_i$ remain arbitrary.
We show that the model admits consistent formulation on the
anti-DeSitter  space.
The  solutions of the system are the massless
irreducible representations of Poincar\'e group
with $N$  nonzero helicities, which are equal to each other.
\end{abstract}
\setcounter{equation}0
\section{The model}
About ten years ago M.Plyushchay proposed the beautiful Lagrangian
  system describing the four-dimensional
massless particle with the helicity $c$
\cite{misha1},
\begin{equation}
{\cal S}=c\int k_1 d{\tilde s}, \quad d{\tilde s}=|{d{\bf x}}|\neq 0,
\end{equation}
where $k_1$ denotes the first curvature of a world-line,
assumed to be  a non-isotropic curve.\\
Quantization  of this model yields the massless irreducible representation
of Poincar\'e group with any integer \cite{misha1}  and half-integer
\cite{rr2} helicity.
This model possesses  the  gauge  $W_3$  symmetry
 reflected in the gauge equivalence of its
 classical trajectories, which are  {\it space-like plane curves},
i.e. $k_1$ is any, $k_2=0$ \cite{rr}.
Another interesting  property of Plyushchay's model is that it admits
formulation on the AdS space \cite{nesterenko}.

{\it Is it possible to generalize
the Plyushchay's model in a $D>4$ dimensional space-time?}

This question is not so trivial, since in $D>6$ massless particles
are specified by  $[(D-2)/2]$ helicities (the weights of the little
Lorentz group $SO(D-2)$), while the Plyushchay's model in $D>6$
has  only one nonvanishing helicity.
On the other hand, only those massless irreps are conformal invariant,
whose all helicities are equal to each other (being any (half)integers
in an even dimensional space-time, and $0,1/2$ in the odd-dimensional
one)\cite{siegel}.

Systematically investigating the actions
\begin{equation}
 {\cal S} =\int {\cal L}({ k}_1,....,{ k}_N)d{\tilde s},
\label{gactions}\end{equation}
one find that the  systems  describing  massless particles with
fixed helicities are of the form
\begin{equation}
{\cal S}=c\int{k}_N d{ s},   \quad d{\tilde s}=|{d{\bf x}}|\neq 0
\label{action}\end{equation}
where  $k_i$ denote the reparametrization invariants (extrinsic curvatures)
of a world-line \cite{tmp}.\\

We establish the following  interesting
properties of this model (for more detail see \cite{massless}):
\begin{itemize}
\item its classical solutions  are space-like curves which obey the
conditions $k_{N+i}=k_{N-i}$, where $i=1,\ldots N\leq N_0=[(D-2)/2]$,
$k_0\equiv 0$; the values of the first $N$ curvatures remain arbitrary,
 so the model  possesses $N+1$ gauge degrees of freedom forming the
algebra of the $W$-type;
\item it describes a massless particle
whose $N$ helicities are equal to each other
$c_1=c_2=\ldots=c_N=c$, and the remaining ones vanish;
\item  it admits a consistent formulation on the
Anti-De Sitter spaces.
\end{itemize}
 \section{The analysis}
In order to obtain  the Hamiltonian formulation
 of the system ,
we should replace
 the action (\ref{action})
 by the classically equivalent one,
whose Lagrangian  depends on the first-order derivatives,
 and then perform the Legendre
transformation.
For this purpose, it is convenient  to use
the recurrent equations for extrinsic curvatures,
which follow from the Frenet equations
for a moving frame $\{{\bf e}_a\}$ of nonisotropic curves:
\begin{eqnarray}
&{\dot{\bf x}}=s{\bf e}_1,\quad {\dot{\bf e}}_a=sK_a^{\;b}{\bf e}_b,
\quad{\bf e}_a{\bf e}_b=\eta_{ab},
\quad a, b=1,\ldots, D&\\
& K_{ab}^{\;.}+K_{ba}^{\;.}=0,\quad
K_{ab}^{\;.}=
\left\{\begin{array}{ccc}
\pm k_a,&{\rm if} &b=a\pm 1\\
0,&{\rm if}&b\neq a\pm 1
\end{array}\right.,\quad k_a \geq 0.& \nonumber
\end{eqnarray}
The Frenet equations
in the Euclidean space
read
\begin{equation}
 {\bf{\dot x}}=s{\bf e}_1,\quad
{\bf\dot e}_a=sk_a{\bf e}_{a+1}-sk_{a-1}{\bf e}_{a-1},\quad\quad
{\bf e}_0={\bf e}_{D+1}\equiv 0,\;\;k_0=k_D=0.
\label{ff}\end{equation}
It is easy to verify that for the transition to the
Frenet equations for non-isotropic curves
in the pseudo-Euclidean space, we should make,
 for some index ${\underline a}$, the
 substitution,
\begin{equation}
({\bf e}_{\underline a},\; sk_{\underline a},\;
sk_{{\underline a}-1}, s)\to
({\rm i}{\bf e}_{\underline a},\; {\rm i}sk_{\underline a},\;
{\rm i}sk_{{\underline a}-1},(-{\rm i})^{\delta_{1{\underline a}}}s).
\end{equation}
 The choice ${\underline a}=1$
means the transition to a time-like curve,
while ${\underline a}=2,\ldots, D$- to space-like ones.
Thus,  throughout the paper we  use the
Euclidean signature.

Taking into account the   expressions (\ref{ff})
 one can replace the initial Lagrangian
(in arbitrary time parametrization $d{\tilde s}={s}d\tau$,
${ s}=|{\bf\dot x}|$) by the following one
\begin{eqnarray}
{\cal  L}&=c\sqrt{{\dot{\bf e}}^2_{N} -(sk)^2_{N-1}} +
{\bf p}({\bf{\dot x}}-{ s}{\bf e}_1)
+\sum_{i=1}^{N-1}{\bf p}_{i}
({\dot{\bf e}}_{i}- { s}k_{i}{\bf e}_{i+1}+
{ s}k_{i-1}{\bf e}_{i-1})
&\nonumber\\
&-{ s}\sum_{i,j=1}^N d_{ij}\left({\bf e}_i{\bf e}_j-\delta_{ij}\right)
&\label{lfo}\end{eqnarray}
where  ${ s}, k_{i}, d_{ij}, {\bf p}_{i-1}, {\bf e}_i$
are independent variables, $k_0=0,{\bf p}_0={\bf e}_0=0$.

Performing the Legendre transformation for
this Lagrangian (see, for  details, \cite{tmp}),
 one gets the
 Hamiltonian system
\begin{equation}
\begin{array}{c}
\omega={\rm d}{\bf p}\wedge {\rm d}{\bf x}+
\sum_{i=1}^N {\rm d}{\bf p}_i\wedge {\rm d}{\bf e}_i,\\
{\cal H}={ s}\left[{\bf p}{\bf e}_1+
\sum_{i=1}^{N}{ k}_{i-1}{\phi}_{i-1.i}
+\frac{{ k}_{N}}{2c}({\Phi}_{N.N}-c^2)
+ \sum_{i,j=1}^Nd_{ij}({\bf e}_i{\bf e}_j-\delta_{ij})\right],
\end{array}
\label{ss}\end{equation}
with the  primary  constraints
\begin{eqnarray}
& {\bf e}_i{\bf e}_j-\delta_{ij}\approx 0,&\label{u}\\
&{\bf p}_N{\bf e}_N\approx 0,\;\;{\bf p}_N{\bf e}_{i-2}\approx 0,
&\label{gauge} \\
&{\bf p}{\bf e}_1 \approx 0,&\label{phi0}  \\
& {\phi}_{i-1.i}\equiv {\bf p}_{i-1}{\bf e}_i
- {\bf p}_{i}{\bf e}_{i-1}\approx 0,\quad
{\Phi}_{N.N}\equiv{\bf p}^2_N-\sum_{i=1}^N({\bf p}_N{\bf e}_i)^2\approx
c^2.&\label{primarylinear}
\end{eqnarray}
It is convenient to introduce  the new variables,
instead of ${\bf p}_i$,
\begin{equation}
\begin{array}{c}
{{\bf p}^\bot}_i\equiv{\bf p}_i-
\sum_{j=1}^N({\bf p}_i{\bf e}_j){\bf e}_j, \quad
 {\bf p}^\bot_i{\bf p}^\bot_j\equiv {\Phi}_{i.j},\quad
{{\bf p}^\bot}_i{\bf e}_j=0,\\
{\phi}_{i.j}\equiv{\bf p}_i{\bf e}_j- {\bf p}_j{\bf e}_i,\\
\chi_{ij}={\bf p}_i{\bf e}_j,\quad i\geq j.
\end{array}
\label{pbot}\end{equation}
Since the constraints (\ref{u})  are conjugated to $\chi_{ij}$
and commute with  ${\bf p}^\bot_i$ and $\phi_{ij}$,
we  impose, without loss of generality,
the gauge conditions $\chi_{ij}\approx 0$ fixing the values
 of $d_{ij}$
\begin{equation}
\chi_{ij}\approx 0:\;\;\Rightarrow \quad
2d_{i.j}= \delta_{ij}\delta_{iN}{k}_Nc.
\label{d}\end{equation}
Notice that in this formulation
 ${ s}$ and ${s}k_i$  play the role of  Lagrangian multipliers,
 so the primary constraints produce either secondary ones,
 or  the explicit relations on the first $N$ curvatures.

The primary constraints
(\ref{phi0}),(\ref{primarylinear})
produce the following set of constraints, which are of the first class
\begin{eqnarray}
&{\bf p}{\bf e}_i\approx 0,\quad
{\bf p}{\bf p}_i\approx 0 ,
\quad  {\bf p}^2\approx 0,&\label{massless} \\
&{\phi}_{ij}\approx 0,\quad
{\Phi}_{ij}-c^2\delta_{ij}\approx 0&,
\label{isospinmax}
\end{eqnarray}
so, the  dimension of  phase space is
\begin{equation}
{\cal D}^{phys}=  2(D-1)+ N(2D-3N-5).
\end{equation}
From the expressions (\ref{massless}) which provide the
model with the  mass-shell and transversality conditions, we conclude
that {\it the nontrivial classical solutions of the system
 (\ref{action})  are the space-like curves  with} $N\leq N_0=[(D-2)/2]$.

Comparing the equations of motion with the Frenet formulae (\ref{ff}),
one finds that
the space-like vectors $({\bf e}_i,
{\bf p}^\bot_i/c\equiv{\bf e}_{2N+1-i})$
define the first $2N$  elements  of
moving frame, while ${\bf p}$ defines  its $(2N+1)-$th, isotropic, element.
We also  get the following  relations on curvatures
\begin{equation}
k_{N+i}=k_{N-i}, \quad i=1,\ldots, N,\quad k_{0}\equiv 0.
\label{curvcurve}
\end{equation}
The first $N$ curvatures remain arbitrary,
{\it hence,  the system possesses $N+1$ gauge degrees of freedom.}\\

Let  introduce the complex variables
$$ {\bf z}_i=\frac{{\bf p}_i+{\rm i} {|c|}{\bf e}_i}{\sqrt{2|c|}}$$
 in which  the Hamiltonian system reads
\begin{equation}
\begin{array}{c}
\omega={\rm d}{\bf p}\wedge {\rm d}{\bf x}+{\rm i}
\sum_{i}{\rm d}{\bf z}_i\wedge {\rm d}{\bar{\bf z}}_i,\\
\;\;\\
{\cal H}=\frac{s}{2c}\left[
{\rm i}{\sqrt{2}}{\bf p}({\bf{\bar z}}_1-{\bf{z}}_1)
+{\rm i}\sum_{i=1}^{N-1}{ k}_i({\bf z}_{i}{\bf{ \bar z}}_{i +1}
-{\bf z}_{i+1}{\bf{ \bar z}}_{i})+
{ k}_N({\bf z}_{N}{\bf{\bar z}}_{N}-c^2)\right],
\end{array}
\end{equation}
while the full set of constraints takes the form
\begin{equation}
\begin{array}{c}
{\bf z}_i{\bf{\bar z}}_j - |c|\delta_{ij}\approx 0, \\
{\bf z}_i{\bf z}_j\approx 0,\\
{\bf p}{\bf z}_i\approx 0, \\
{\bf p}^2\approx 0.
\end{array}
\label{conscomp} \end{equation}
In these coordinates
the equations of motion become holomorphic one
\begin{equation}
\begin{array}{c}
\dot{\bf x}={\rm i}\left({\bf z}_1-{\bf{\bar z}}_1\right),\\
\dot{\bf z}_{i-1}=-{\rm i}\delta_{1.i-1}{\bf p}
+k_{i-1}{\bf z}_{i}-k_{i-2}{\bf z}_{i-2},\\
\dot{\bf z}_{N}=-{\rm i}\delta_{1.N}{\bf p}
+{\rm i}k_{N}{\bf z}_{N}-k_{N-1}{\bf z}_{N-1},\\
{\dot{\bf p}}=0,
\end{array}
\end{equation}
which can be considered as the
implicit indicator of the $W-$algebraic origin of a
system's gauge symmetries \cite{gm}.
Another argument is that the equations of motion for ${\bf z}_i$
can be rewritten in the form\cite{radul}
$$
{\hat  L}{\bf z}_1=
\sum_{i=0}^N\lambda_i(k_{i(\alpha)})
\frac{{\rm d}^i{\bf z}_1}{{{\rm d}\tau}^i}=0.
$$
 This is in correspondence
with the claim  that the gauge symmetries of
the action (\ref{action}) define the classical limit of $W_{N+2}$
 algebra (which was proved for $N=1,2$)\cite{rr2}.\\

Note that the rotation generators can be represented as follows
\begin{equation}
             {\bf M}={\bf p}\times {\bf x}+
\frac{{\rm i}}{c}\sum_{i=1}^N{\bf z}_i\times{\bf{\bar z}}_i.
\end{equation}
So, the system has
 the following helicities
(the weights of the little Lorentz group $SO(D-2)$)
\begin{equation}
c_1=\ldots=c_N =c\neq 0,\quad c_{N+1}=\ldots=c_{N_0}=0.
\end{equation}
To quantize the system, we should choose the polarization
\begin{equation}
 {\hat{\bf{\bar z}}}_i=\frac{\partial}{\partial {\bf z}_i}, \quad
{\hat{\bf x}}=\frac{\partial}{\partial {\bf p}},\quad\quad
 {\hat{\bf{ z}}}_i={\bf z}_i, \quad{\hat{\bf p}}={\bf p}.
\end{equation}
Then,  after standard manipulations
we get that the wave function is of the form
\begin{equation}
  \Psi({\bf p}, {\bf z}_i)=\psi({\bf p})_{A(1)A(2)\ldots A(N)}
z^{A(1)}_1 z^{A(2)}_2\ldots z^{A(N)}_N,
\end{equation}
where
$A(i)\equiv A^{(i)}_1\ldots A^{(i)}_{|c|}$,
$z^{A(i)}_i\equiv z^{A^{(i)}_1}_i\cdots z^{A^{(i)}_{|c|}}_i$,
$|c|=1,\;2,\;3,\ldots\;\;$.

Here  $\psi(\bf p)_{\ldots}$ is  the tensor of  $cN$-th rank,
whose symmetries are given by the $N\times c$ Young tableau
\begin{center}
\begin{tabular}{|c|c|c|}
\hline
$A^1_1$  & $\cdots\cdots\cdots$ & $A^1_{|c|}$\\
\hline
$A^2_1$  & $\cdots\cdots\cdots$ & $A^2_{|c|}$  \\
\hline
$\vdots$ & $\quad\quad\quad$ & $\vdots$    \\
\hline
$A^N_1$  & $\cdots\cdots\cdots$ & $A^N_{|c|}$     \\
\hline
\end{tabular}\quad\quad .
\end{center}
In addition, the tensor  $\psi({\bf p})_{\ldots}$
should satisfy  the transversality and mass-shell conditions:
\begin{equation}
\begin{array}{c}
   {\bf p}^2 \psi({\bf p})_{A(1)B(2)\ldots C(N)}=0\quad
p^A\psi({\bf p})_{\ldots A\ldots}=0.
\end{array}
\end{equation}
Due to Siegel \cite{siegel}, the (massless) irreducible
representations  of the system possess conformal invariance, iff
$N=N_0$, where $c_1=\ldots=c_{N_0}$ is  (half)integer
for even $D$ and $c_N=0,1/2$  for odd $D$.
Hence, the model under
consideration  possesses conformal symmetry if
$D=2p,\quad N=[(D-2)/2],$ since the helicities are integers.\\

Let us reformulate our  model on the spaces with constant curvature.
On the curved spaces the Frenet equations read
\begin{equation}
 \frac{{\rm d}{\bf x}}{{ s}{\rm d}\tau}={\bf e}_1,\quad
\frac{{\rm d}{\bf e}_a}{{ s}{\rm d}\tau}=k_a{\bf e}_{a+1}
-k_{a-1}{\bf e}_{a-1},
\end{equation}
where
$$
\frac{D}{{\rm d}\tau}\equiv\frac{{\rm d}}{{\rm d}\tau}
+{\hat\Gamma}({{\bf \dot x}}),
 \quad{\bf e}_a{\bf e}_b=\delta_{ab},\quad
{\bf e}_0={\bf e}_{D+1}\equiv 0,
$$
 while $({\hat \Gamma})^A_B\equiv\Gamma^A_{BC}{\dot x}^C$,
$\Gamma^A_{BC}$ are the Christoffel symbols of
the  metric $g_{AB}(x)$ of underlying manifold.

Performing  the manipulations, similar to the flat case,
we get the Hamiltonian system
\begin{equation}
\begin{array}{c}
\omega={\rm d}{\bf p}\wedge {\rm d}{\bf x}+
\sum_{i=1}^N {\rm d}{\bf p}_i\wedge {\rm d}{\bf e}_i\\
{\cal H}={\tilde s}\left[{\bf \pi}{\bf e}_1+
\sum_{i=1}^{N}{ k}_{i-1}{\phi}_{i-1.i}
+\frac{{ k}_{N}}{2c}({\Phi}_{N.N}-c^2)
+ \sum_{i,j=1}^Nd_{ij}({\bf e}_i{\bf e}_j-\delta_{ij})\right],
\end{array}
\label{ssgr}\end{equation}
whose primary constraints are given by the expressions
\begin{equation}
\begin{array}{cc}
{\bf \pi}{\bf e}_1 \approx 0,&\quad\\
 {\bf e}_i{\bf e}_j-\delta_{ij}\approx 0,&\quad\\
{\bf p}_N{\bf e}_N\approx 0,\;\;{\bf p}_N{\bf e}_{i-2}\approx 0,
&\quad \\
 {\phi}_{i-1.i}\equiv {\bf p}_{i-1}{\bf e}_i
- {\bf p}_{i}{\bf e}_{i-1}\approx 0,&\quad\\
{\Phi}_{N.N}\equiv{\bf p}^2_N-\sum_{i=1}^N({\bf p}_N{\bf e}_i)^2\approx
c^2.&\quad
\end{array}
\label{primarylineargr}
\end{equation}
 where ${\bf\pi}\equiv
{\bf p}-{\bf \Gamma},\quad
 \Gamma_A\equiv
\sum_{i=1}^N{\Gamma}^C_{AB}{p}_{(i)C}{e}^B_i$.\\
It is easy to see that the Poisson brackets of the functions
 $\phi_{ij}, \Phi_{ij}$ remain unchanged, as well as
those
with ${\bf\pi}{\bf a}$, where ${\bf a}={\bf e}_i, {\bf p}_i$, ${\bf \pi }$.
On the other hand,
$$
\{{\bf\pi a},{\bf\pi b}\}=
\sum_{i=1}^N R({\bf p}_i|{\bf e}_i,{\bf a},{\bf b}),
$$
where $R(\;,\;,\;,\;)$ is the curvature tensor of underlying manifold.

Taking into account the expression for the Riemann tensor of
 the  constant curvature spaces,
we conclude that the constraint algebra of the system (\ref{ssgr})
is isomorphic to the one on the constant curvature spaces (\ref{ss}).

Therefore,  the model (\ref{action}) admits the consistent
 formulation on the
anti-De Sitter spaces.

{\large Acknowledgments.}
The author is  grateful to  S. Lyakhovich, I.V.Tyutin and  M. Vasiliev
for valuable discussions and comments,
and O. Khudaverdian, R.Metsaev, C. Sochichiu for  interest in
the work and useful remarks.

\end{document}